# Heat and Mass Transfer during Chemical Vapor Deposition on the Particle Surface Subjected to Nanosecond Laser Heating

## Quan Peng[1], Yuwen Zhang[1,2], Yaling He[1], Yijin Mao[2]

1: Key Laboratory of Thermal Fluid Science and Engineering of MOE, School of Energy and Power Engineering, Xi'an Jiaotong University, Xi'an, China, winnie_0805_pooh@163.com (Q. Peng), yalinghe@mail.xjtu.edu.cn (Y.L. He)
2: Department of Mechanical and Aerospace Engineering, University of Missouri, Columbia, USA, zhangyu@missouri.edu (Y.W. Zhang), ymk89@mail.missouri.edu (Y.J. Mao)

**Abstract** A thermal model of chemical vapor deposition of titanium nitride (TiN) on the spherical particle surface under irradiation by a nanosecond laser pulse is presented in this paper. Heat and mass transfer on a single spherical metal powder particle surface subjected to temporal Gaussian heat flux is investigated analytically. The chemical reaction on the particle surface and the mass transfer in the gas phase are also considered. The surface temperature, thermal penetration depth, and deposited film thickness under different laser fluence, pulse width, initial particle temperature, and particle radius are investigated. The effect of total pressure in the reaction chamber on deposition rate is studied as well. The particle-level model presented in this paper is an important step toward development of multiscale model of LCVI.

## 1. Introduction

Gas-phase Solid Freeform Fabrication (SFF) with Laser Chemical Vapor Deposition (LCVD) or Laser Chemical Vapor Infiltration (LCVI) is an emerging manufacturing technology that can directly produce three dimensional (3-D) parts from CAD data (Conley and Marcus, 1997). The LCVD/LCVI technique, which based on reactions initiated pyrolytically, photolytically or a combination of both (Marcus et al., 1993), can build functional structure via deposition of solid materials from gas precursors irradiated by a laser beam in an environmentally controlled chamber. In the pyrolytically LCVD/LCVI process, a laser beam locally heats the substrate to create a hot spot where a thermally activated chemical reaction is induced; the resulting product, which is a thin film of material, deposits on the substrate due to chemisorption. In the photolytically LCVD/LCVI process, the laser is tuned to an electric or vibrational level of the gas; the irradiated gaseous substance decomposes and the products stick to the substrate surface to form the thin film. During these two processes, lines can be formed by multiple laser scans and are subsequently interwoven to form a part layer. Consecutive layers can be deposited to create a three-dimensional part according to the CAD design. The most recent LCVD/LCVI advances as reported in various journals and proceedings are well documented by Zhang (2010).

While the LCVD technique uses precursors to directly create free-standing part, the LCVI, on the other hand, uses gas precursors and powder particles to build three-dimensional parts, which is similar to other powder-based SFF techniques, such as Selective Laser Sintering (SLS). In the LCVI process, powder particles are bounded together through deposition of solid material on the particle surface by decomposition of gas precursors. During the LCVI process, a thin powder layer (100-250 μm thick) is scanned by laser to form the two-dimensional slice to an underlying solid piece, which consists of a series of stacked two-dimensional slices. A fresh powder layer is spread after laser scanning and the process is repeated. Loose powder is removed after the part is extracted from its bin, and the finished part has a composite structure consisting of starting powder bonded into a matrix of the deposited materials (Birmingham and Marcus, 1994). The advantages of LCVI over LCVD include (a) uninfiltrated powder provides support for producing overhangs, (b) confining the deposition to thin powder layers provide dimensional control in the direction of growth, and (c) it is possible to tailor local chemistry and microstructures (Jakubenas et al., 1997). Much experimental work on LCVI has been conducted and various applications of this new technique have been discovered. Sun et al. (1998) fabricated in situ thermocouples into micro-components using LCVI to build the bulk shape. Crocker et al. (2000; 2001; 2002) presented LCVI of SiC into metal and ceramic powders to fabricate composite parts.

Accurate prediction and control of LCVI require a thermal model of the process. The physical and chemical phenomena occurred in LCVI are similar to that in LCVD, which has been investigated extensively and detailed works are presented by Kar et al. (1991), Conde et al. (1992), Zhang and Faghri (2000), Zhang (2003) and Lee et al. (1995). The only difference is that deposition occurs on the powder particle surface during LCVI, instead of on the top of non-porous flat substrate in LCVD. Laser processing in LCVI is complicated since the irradiated material responds differently from that in the case for a simple, fully dense material. Thermal modeling of the LCVI process requires knowledge about heat and mass transfer in the precursors and the porous powder bed, and chemical reaction on the particle surface. Convective heat transfer in porous material has been extensively investigated in the past and detailed reviews are presented by Nield and Bejan (1999) and Kaviany (1995). In addition, radiation heat transfer in the powder bed also plays an important role in thermal analysis due to the high temperature involved. The distinctive feature of heat transfer in the LCVI process is that the porosity is not constant. During the entire process, the porosity may change from a value up to 0.6 to nearly zero.

Dai et al. (2000) presented numerical simulation of LCVI using a finite element commercial code ANSYS. The effect of variational porosity on the powder bed properties was taken into account. The density of the powder bed was directly correlated to the temperature and thus the chemical reaction model was not necessary. Dai et al. (2001) introduced a finite element model with the same method and the laser power was modified from one time step to the next to ensure that the powder bed temperature was constant. The models presented by Dai et al. (2000; 2001) were improved later through incorporating a densification model by infiltration based on experimental growth rate (Dai et al., 2003).

Compared to LCVD, the efforts on simulation of LCVI are extremely limited at this time. Moreover, most of the existing works on LCVI were conducted on the powder bed level, and mass transfer in the precursors and powder bed was not taking into account in any of them. In this paper, a thermal model will be developed to investigate heat and mass transfer during chemical vapor deposition on the surface of a single powder particle irradiated by a nanosecond laser pulse. The particle surface temperature distribution, thermal penetration depth, and the thickness of deposited film will be obtained through numerical simulation. In addition, the effects of laser fluence and pulse width, particle radius, initial particle temperature, and total pressure in the reaction chamber on the simulation results will also be investigated.

## 2. Physical model

During the LCVI process, the diameter of the particle is much smaller than that of the laser beam, which is in turn much smaller than the dimension of the final part. Since the laser irradiates the powder bed from a distance of several powder-sphere diameters away, a nearly homogeneous distribution of the heat flux within the penetrated layer can be assumed due to multiple scattering of the radiation (Fischer et al., 2002). The physical model of LCVI on a single powder particle surface subjected to temporal Gaussian heat flux from a nanosecond pulsed laser beam is illustrated in Fig. 1; a metal powder particle made of Incoloy 800 is surrounded by a mixture of $H_2$, $N_2$, and $TiCl_4$. Because of symmetry of the spherical particle as well as the assumption of uniform heat flux distribution, the problem can be simplified to be one-dimensional in the *r*-direction. The origin of time is chosen to be when the heat flux is at its maximum, thus the time-dependent heat flux is expressed as

$$q''(t) = q_0'' e^{-\ln 2 \frac{t^2}{t_p^2}} \quad (1)$$

where $q_0''$ is the maximum heat flux, and $t_p$ is the half-width of the laser pulse at half maximum.

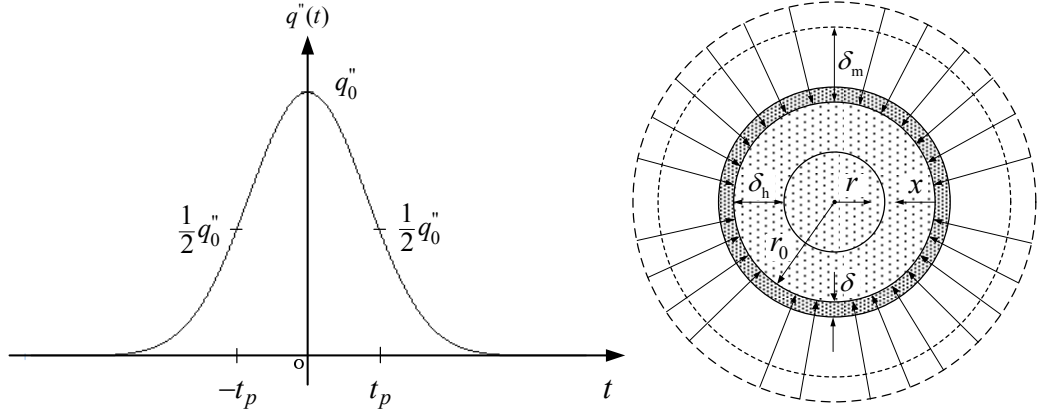

**Fig. 1.** The physical model.

After the onset of the laser pulse, the particle surface temperature raises rapidly due to sensible heat absorption. The vapor deposition starts when the surface temperature reaches the chemical reaction threshold temperature of 1173 K, and the concentration difference resulting from chemical reaction on the particle surface becomes the driving force of mass transfer. The surface heat flux reaches its peak value at the time $t = 0$, after which the heat flux will decrease. The deposition process finishes when the surface temperature decreases to the threshold temperature. The laser energy absorbed by the particle surface will be distributed to two parts during the chemical reaction: the first part goes into the powder grain through conduction, and the second part is consumed by the chemical reaction (Zhang and Faghri, 2000). Thus the physical model of the LCVI process includes: heat transfer in the powder particle, chemical reaction on the particle surface, and mass transfer in the gases. After the pulse is off, the internal temperature distribution of the particle will be uniform well before the next laser pulse is launched. Consequently, the whole process can be modeled as a recurrence of the two stages by using the uniform temperature from the previous pulse as the initial condition in the preheating stage of the next pulse (Konrad et al., 2007). All physical properties of the materials, such as density, conductivity and so on, are independent of temperature and pressure during the whole process.

*2.1. Governing equations for heat transfer in the powder particle*

The heat transfer within the powder particle can be described as a pure conduction problem. The heat conduction equation within the particle is

$$\frac{\partial \theta}{\partial t} = \frac{\alpha}{(r_0 - x)^2} \frac{\partial}{\partial x}[(r_0 - x)^2 \frac{\partial \theta}{\partial x}],$$
$$0 \le x \le r_0, \; t > -\infty \quad (2)$$

which is subject to the following initial and boundary conditions:

$$\theta = 0, \quad 0 \le x \le r_0, \; t \to -\infty \quad (3)$$

$$\frac{\partial \theta}{\partial x} = 0, \quad x = r_0, \; t > -\infty \quad (4)$$

$$-k\frac{\partial \theta}{\partial x} = q''(t), \quad x = 0, \; t > -\infty \quad (5)$$

*2.2. Chemical reaction on the particle surface*

In the chemical reaction undergone on the particle surface, TiN is produced according to the overall chemical reaction expressed as (Conde et al., 1992)

$$\mathrm{TiCl_4}(g) + 2\mathrm{H_2}(g) + \frac{1}{2}\mathrm{N_2}(g) \rightarrow \mathrm{TiN}(s) + 4\mathrm{HCl}(g)$$
$$(6)$$

The production rate of TiN is defined as the mass of TiN produced per unit area per unit time, and can be calculated by

$$\dot{m}_{TiN} = K(T)C_{H_2}C_{N_2}^{1/2}C_{TiCl_4} \quad (7)$$

where $K(T)$ is the reaction rate constant and can be calculated by Arrhenius equation (Zhang and Faghri, 2000)

$$K(T) = K_0' \exp(-\frac{E}{R_u T_s}) \quad (8a)$$

Since the mole fractions of $H_2$ and $N_2$ inside the reaction chamber are at least one order of magnitude higher than that of $TiCl_4$ (Conde et al., 1990), the variation of $C_{H_2}$ and $C_{N_2}$ in the deposition process can be neglected, i.e., $C_{H_2}$ and $C_{N_2}$ can be treated as constants in Eq. (7). Thus, another variable $K_0$ is introduced

$$K_0 = (C_{H_2})_i (C_{N_2})_i^{1/2} K_0' \quad (8b)$$

and Eq. (7) is rewritten as:

$$\dot{m}_{TiN} = K_0 \exp(-\frac{E}{R_u T_s})C_{TiCl_4} \quad (9)$$

Thus, the rate that $TiCl_4$ is consumed becomes:

$$\dot{m}_{TiCl_4} = K_0 \exp(-\frac{E}{R_u T_s})C_s \cdot \frac{M_{TiCl_4}}{M_{TiN}} \quad (10)$$

where $C_s$ represents the concentration of $TiCl_4$ at the particle surface.

The deposit rate is expressed as

$$\frac{d\delta}{dt} = \gamma_{TiN}\frac{\dot{m}_{TiN}}{\rho_{TiN}} = \gamma_{TiN}\frac{K_0}{\rho_{TiN}}\exp(-\frac{E}{R_u T_s})C_s \quad (11)$$

The sticking coefficient $\gamma_{TiN}$ is defined as

$$\gamma_{TiN} = \begin{cases} 1 & T_s < T_{min} \\ 1+\frac{T_{min}-T_s}{T_{max}-T_{min}} & T_{min} \leq T_s \leq T_{max} \\ 0 & T_s > T_{max} \end{cases} \quad (12)$$

where $T_{min}$ is a threshold temperature below which the product of the chemical reaction can be fully stuck on the particle surface, and $T_{max}$ is another threshold temperature above which no product of chemical reaction can stick to the surface. If the surface temperature is between $T_{min}$ and $T_{max}$, the product of chemical reaction can only be partially stuck on the surface. The values of $T_{min}$ and $T_{max}$ are chosen as 1273 K and 1673 K, respectively.

*2.3. Governing equations for mass transfer in the gases*

From Eq. (11), it can be seen that the concentration of $TiCl_4$ must be solved in order to predict the deposition rate. The concentration of $TiCl_4$ is governed by the following mass diffusion equation

$$\frac{\partial C}{\partial t} = D\frac{1}{r^2}\frac{\partial}{\partial r}(r^2\frac{\partial C}{\partial r}), \quad r > r_0 \quad (13)$$

which is subject to the following initial and boundary conditions:

$$C = C_i, \quad t = -\infty \quad (14)$$

$$C = C_i, \quad r = \infty \quad (15)$$

$$-\dot{m}_{TiCl_4} = -D\frac{\partial C}{\partial r}\Big|_{r=r_0} \quad (16)$$

## 3. Integral approximate solution

*3.1. Solution for heat transfer in the powder particle*

*3.1.1. Solution for the t < 0 stage during one laser pulse*

When the particle surface is exposed to heat flux, the heat will penetrate the surface and begin to conduct inward. The depth which the heat flux has penetrated to is called the thermal penetration depth, $\delta_h$, beyond which the temperature remains to be the initial temperature. Therefore, the following two boundary conditions are valid

$$\theta(x,t) = 0, \quad x \geq \delta_h(t), t > -\infty \quad (17)$$

$$\frac{\partial \theta}{\partial x} = 0, \quad x \geq \delta_h(t), t > -\infty \quad (18)$$

As the heat flux penetrates the particle surface, the thermal penetration depth, $\delta_h$, will increase. It is assumed that the thermal penetration depth will not reach the center during this stage since the heat flux is high. It will be necessary to monitor the temperature distribution of the powder particle so that the onset of chemical reaction can be detected.

Integrating both sides of Eq. (2) with respect to $x$ in the interval of $(0, \delta_h)$ and applying Eqs. (5), (17) and (18), the integral energy equation becomes

$$\frac{\partial}{\partial t}\int_0^{\delta_h}(r_0-x)^2\theta(x,t)dx = \frac{\alpha q_0'' r_0^2}{k}e^{-\ln 2\frac{t^2}{t_p^2}} \quad (19)$$

Assuming that the temperature distribution can be approximated by a second degree polynomial function, $\theta = A + Bx + Cx^2$, and solving the unknown constants using the boundary conditions of Eqs. (5), (17) and (18), the temperature distribution in the thermal penetration depth becomes

$$\theta(x,t) = \frac{q_0''(\delta_h-x)^2}{2k\delta_h}e^{-\ln 2\frac{t^2}{t_p^2}}, \quad \delta_h < r_0, \ t < 0 \quad (20)$$

Substituting Eq. (20) into Eq. (19) yields

$$\frac{\partial}{\partial t}[\frac{1}{6}\delta_h^2 e^{-\ln 2\frac{t^2}{t_p^2}}(r_0^2 - \frac{1}{2}r_0\delta_h + \frac{1}{10}\delta_h^2)] = \alpha r_0^2 e^{-\ln 2\frac{t^2}{t_p^2}} \quad (21)$$

which is subject to the initial condition $\delta_h = 0$, when $t = -\infty$.

Integrating Eq. (21) with respect to $t$ in the interval $(-\infty, t)$ one obtains

$$\frac{1}{60}\delta_h^4 e^{-\ln 2\frac{t^2}{t_p^2}} - \frac{1}{12}\delta_h^3 r_0 e^{-\ln 2\frac{t^2}{t_p^2}} + \frac{1}{6}\delta_h^2 r_0^2 e^{-\ln 2\frac{t^2}{t_p^2}} -$$

$$\frac{\alpha r_0^2 t_p \sqrt{\pi}}{2\sqrt{\ln 2}}[1+erf(\frac{\sqrt{\ln 2}}{t_p}t)] = 0, \quad \delta_h < r_0, \ t < 0 \quad (22)$$

The physically reasonable solution of $\delta_h$ will be the one, which is positive and real, among all four roots of Eq. (22).

*3.1.2. Solution for the t > 0 stage during one laser pulse*

When $t > 0$, the surface heat flux begins to decrease, and the initial condition for this stage can be obtained from the continuity of the temperature profile within the particle at the time $t = 0$ requires that

$$\theta(x,t=0^+) = \theta(x,t=0^-) = \frac{q_0''(\delta_h-x)^2}{2k\delta_h}, \quad \delta_h < r_0 \quad (23)$$

The integral approximate solution cannot obtain good results for the case with decreasing heat flux, thus we divide the problem into two sub-problems:

$$\theta = \theta_1 + \theta_2 \quad (24)$$

where both $\theta_1$ and $\theta_2$ satisfy Eq. (2). The boundary and initial conditions of $\theta_1$ are

$$-k\frac{\partial \theta_1}{\partial x} = q_1''(t) = q_0'', \quad x = 0, \ t > 0 \quad (25)$$

$$\theta_1(x,0) = \theta(x,0) = \frac{q_0''(\delta_h-x)^2}{2k\delta_h}, \quad t = 0, \ \delta_h < r_0 \quad (26)$$

$$\delta_{h1}(t) = \delta_h, \quad t = 0 \quad (27)$$

$$\theta_1 = 0, \quad x \geq \delta_{h1}(t), \ t > 0 \quad (28)$$

$$\frac{\partial \theta_1}{\partial x} = 0, \quad x \geq \delta_{h1}(t), \ t > 0 \quad (29)$$

and the initial and boundary conditions of $\theta_2$ are

$$-k\frac{\partial \theta_2}{\partial x} = q_2''(t) = -q_0'' + q_0''e^{-\ln 2\frac{t^2}{t_p^2}}, \quad x = 0, \ t > 0 \quad (30)$$

$$\theta_2(x,0) = 0, \quad t = 0, \ 0 \leq x \leq r_0 \quad (31)$$

$$\delta_{h2}(t) = 0, \quad t = 0 \quad (32)$$

$$\theta_2 = 0, \quad x \geq \delta_{h2}(t), \ t > 0 \quad (33)$$

$$\frac{\partial \theta_2}{\partial x} = 0, \quad x \geq \delta_{h2}(t), \ t > 0 \quad (34)$$

*3.1.2.1. Solution of sub-problem for $\theta_1$*

The solution of sub-problem for $\theta_1$ is independent of that for $\theta_2$ and can be obtained using the integral approximate method. Following a procedure similar to that of the $t < 0$ stage, the solution of the $\theta_1$ sub-problem for $\delta_{h1} < r_0$ is

$$\theta_1(x,t) = \frac{q_0''(\delta_{h1}-x)^2}{2k\delta_{h1}}, \quad \delta_{h1} < r_0, \ t > 0 \quad (35)$$

where $\delta_{h1}$ can be obtained from

$$\frac{1}{60}\delta_{h1}^4 - \frac{1}{12}r_0\delta_{h1}^3 + \frac{1}{6}r_0^2\delta_{h1}^2 - \frac{1}{60}\delta_h^4 + \frac{1}{12}r_0\delta_h^3 -$$

$$\frac{1}{6}r_0^2\delta_h^2 = \alpha r_0^2 t, \quad \delta_{h1} < r_0, \ t > 0 \quad (36)$$

Only one of the four roots of Eq. (36) will be both real and positive, and this will be the value of $\delta_{h1}$ at each time step before $\delta_{h1}$ reaches the center, $r_0$.

After the thermal penetration depth, $\delta_{h1}$, reaches the center of the particle, the boundary conditions of Eqs. (28) and (29) will no longer be useful. Assuming that the temperature profile can be approximated by a second degree polynomial and determining the unknown constants using Eqs. (25) and (26), the temperature distribution in the particle becomes (Dombrovsky and Sazhin, 2003)

$$\theta_1(x,t) = \frac{q_0''(r_0-x)^2}{2kr_0} + T_c(t) - T_i, \quad \delta_{h1} = r_0, \ t > 0 \tag{37}$$

where $T_c(t)$ is the temperature at the center of the particle. Integrating both sides of Eq. (2) with respect to $x$ in the interval of $(0, r_0)$ and applying Eqs. (28) and (29), the integral energy equation becomes

$$\frac{\partial}{\partial t} \int_0^{r_0} (r_0-x)^2 \theta_1(x,t) dx = \frac{\alpha q_0'' r_0^2}{k} \tag{38}$$

Substituting Eq. (37) into Eq. (38) yields

$$\frac{\partial}{\partial t}[\frac{q_0'' r_0^4}{10k} + \frac{1}{3}r_0^3(T_c(t)-T_i)] = \frac{\alpha q_0'' r_0^2}{k} \tag{39}$$

which is subject to the initial condition $T_c(t) = T_i$ when $t = t_{\delta h1}$, where $t_{\delta h1}$ is the time at which $\delta_{h1}$ reaches $r_0$, the center of the particle. Eq. (39) can be integrated with respect to $t$ in the interval $(t_{\delta h1}, t)$ to yield

$$\frac{1}{3}r_0^3(T_c(t)-T_i) = \frac{\alpha q_0'' r_0^2}{k}(t-t_{\delta_{h1}}), \quad \delta_{h1} = r_0, \ t > 0 \tag{40}$$

Solving for $T_c(t)-T_i$ from Eq. (40) and substituting the result into Eq. (37), the temperature distribution becomes

$$\theta_1(x,t) = \frac{q_0''(r_0-x)^2}{2kr_0} + \frac{3\alpha q_0''}{r_0 k}(t-t_{\delta_{h1}}),$$
$$\delta_{h1} = r_0, \ t > 0 \tag{41}$$

3.1.2.2. Solution of sub-problem for $\theta_2$

Since the initial temperature of the $\theta_2$ sub-problem is uniform [see Eq. (31)], its thermal penetration depth, $\delta_{h2}$, will originate from the surface at the time $t = 0$ [see Eq. (32)] and begin to penetrate inward. Before $\delta_{h2}$ reaches the center of the particle, $\theta_2$ can be obtained by following a procedure identical to that of the $t < 0$ stage, the result of which is

$$\theta_2(x,t) = \frac{q_0''(\delta_{h2}-x)^2}{2k\delta_{h2}}(e^{-\ln 2 \frac{t^2}{t_p^2}} - 1), \quad \delta_{h2} < r_0, \ t > 0 \tag{42}$$

where $\delta_{h2}$ can be obtained from

$$\frac{1}{6}\delta_{h2}^2(r_0^2 - \frac{1}{2}r_0\delta_{h2} + \frac{1}{10}\delta_{h2}^2)(e^{-\ln 2 \frac{t^2}{t_p^2}} - 1) =$$
$$\alpha r_0^2[\frac{t_p\sqrt{\pi}}{2\sqrt{\ln 2}}erf(\frac{\sqrt{\ln 2}}{t_p}t) - t], \quad \delta_{h2} < r_0, \ t > 0 \tag{43}$$

Only one of the four roots of this polynomial will be both real and positive and this root will be the value of $\delta_{h2}$ for the particular time.

After $\delta_{h2}$ has reached the center of the particle, $\theta_2$ can be obtained using the integral approximate solution

$$\theta_2(x,t) = \frac{q_0''(r_0-x)^2}{2kr_0}(e^{-\ln 2 \frac{t^2}{t_p^2}} - 1) - \frac{3q_0'' r_0}{10k}(e^{-\ln 2 \frac{t^2}{t_p^2}} -$$
$$e^{-\ln 2 \frac{t_{\delta h2}^2}{t_p^2}}) - \frac{3\alpha q_0''(t-t_{\delta_{h2}})}{kr_0} + \frac{3\alpha q_0'' t_p \sqrt{\pi}}{2kr_0 \sqrt{\ln 2}}[erf(\frac{\sqrt{\ln 2}}{t_p}t)$$
$$-erf(\frac{\sqrt{\ln 2}}{t_p}t_{\delta_{h2}})], \quad \delta_{h2} = r_0, \ t > 0 \tag{44}$$

where $t_{\delta h2}$ is the time at which $\delta_{h2}$ reaches the center of the powder particle.

3.2. Solution for mass transfer in the gases

When the surface temperature reaches the chemical reaction threshold temperature, the vapor deposition starts and the titanium tetrachloride will be consumed at the surface of the particle. As a result, the concentration of titanium tetrachloride near the particle surface will be lowered. Similar to thermal penetration depth, a mass penetration depth, $\delta_m$, beyond which the concentration of TiCl$_4$ is equal to its initial value, can be introduced. The following two boundary conditions at the mass penetration depth are valid

$$\frac{\partial C}{\partial r}\bigg|_{r=r_0+\delta_m} = 0 \tag{45}$$

$$C(r,t)\bigg|_{r=r_0+\delta_m} = C_i \tag{46}$$

Integrating both sides of Eq. (13) with respect to $r$ in the interval of $(r_0, r_0 + \delta_m)$ and applying Eqs. (16), (45) and (46), the integral mass diffusion equation becomes

$$\frac{\partial}{\partial t}\int_{r_0}^{r_0+\delta_m} r^2 C(r,t) dr = (r_0+\delta_m)^2 C_i \cdot \frac{d\delta_m}{dt} - r_0^2 \dot{m}_{TiCl_4} \tag{47}$$

which is subject to the initial condition $\delta_m = 0$ when $t = t_{start}$, where $t_{start}$ is the time at which chemical reaction begins.

Assuming that the concentration distribution of TiCl$_4$ is a second degree polynomial function and invoking boundary conditions of Eqs. (16), (45) and (46) to solve the unknown constants, the concentration distribution of TiCl$_4$ in the mass penetration depth becomes

$$C(r,t) = C_i - \frac{\dot{m}_{TiCl_4}}{2\delta_m D}(r_0 + \delta_m - r)^2, \; r_0 \leq r \leq r_0 + \delta_m \quad (48)$$

Substituting Eq. (48) into Eq. (47) and integrating it with respect to $t$ in the interval ($t_{start}$, $t$), one obtains

$$\delta_m^4 + 5r_0\delta_m^3 + 10r_0^2\delta_m^2 - \frac{60D}{\dot{m}_{TiCl_4}}r_0^2 \cdot \int_{t_{start}}^{t}\dot{m}_{TiCl_4}dt = 0 \quad (49)$$

Again, only the one, which is real and positive, among the four roots of Eq. (49) will be the appropriate value of $\delta_m$ for the particular time.

From Eq. (48) the concentration of TiCl$_4$ at the particle surface can be obtained

$$C_s = C_i - \frac{\dot{m}_{TiCl_4}}{2D}\delta_m \quad (50)$$

Substituting Eq. (50) into Eq. (10), the mass flux of TiCl$_4$ at surface becomes:

$$\dot{m}_{TiCl_4} = \frac{C_i}{\frac{1}{K_0}\frac{M_{TiN}}{M_{TiCl_4}}\exp(\frac{E}{R_u T_s}) + \frac{\delta_m}{2D}} \quad (51)$$

The thickness of the deposited film, $\delta$, can be solved through iteration. The procedure of which for a time step is outlined as follows:

1. Guess a value of the consumption rate of TiCl$_4$, $(\dot{m}_{TiCl_4})_{guess}$, then the integral term in Eq. (49) can be evaluated using trapezoidal rule.
2. Solve for $\delta_m$ from Eq. (49).
3. Calculate the consumption rate of TiCl$_4$, $(\dot{m}_{TiCl_4})_{new}$, from Eq. (51).
4. Compare the consumption rate of TiCl$_4$ obtained in Step 3 with the guessed value in Step 1. If $\left|(\dot{m}_{TiCl_4})_{new} - (\dot{m}_{TiCl_4})_{guess}\right| \leq \varepsilon$, where $\varepsilon$ is a small tolerance value, end the iteration and calculate $\delta$ from Eq. (11) through trapezoidal integration and go to the next time step. If not, update the value of $(\dot{m}_{TiCl_4})_{guess}$ and go back to Step 1.

## 4. Results and discussion

The computer codes were written to obtain the results associated with current model. The total pressure in the chamber is 700 kPa (Wallenberger and Nordine, 1993; 1994). The partial pressures of TiCl$_4$, N$_2$ and H$_2$ are assigned in the ratio 1:14.28:14.28. The initial temperature of the particle and gas is 293 K. The concentrations of different species can be obtained by using ideal gas law. With the conditions specified above, the constant $K_0$ in Eq. (8b) is $1.3295 \times 10^3$ m/s. The activation energy of the chemical reaction is taken to be $E = 51.02$ kJ/mol (Conde et al., 1992). The radius of the laser beam, which is defined as the radius where the laser intensity is $1/e^2$ of the intensity at the center of the laser beam, is $1.0 \times 10^{-3}$ m. The absorptivity of the laser beam at the particle surface is taken to be 0.23 (Conde et al., 1992). The heat flux at the particle surface, $q''$ (W/m$^2$) is related to the total energy exerted on the particle by one pulse, or the laser fluence $J$ (J/m$^2$):

$$J = \int_{-\infty}^{\infty} q''(t)dt = \int_{-\infty}^{\infty} q_0'' e^{-\ln 2 \frac{t^2}{t_p^2}}dt = q_0'' t_p \frac{2\sqrt{\pi}}{\sqrt{\ln 2}} \quad (52)$$

Chemical vapor deposition process can be investigated by changing simulation parameters.

Figure 2 shows the surface temperature of the powder particle for nine laser pulses. The laser pulse frequency of 5000 Hz corresponds to 200 μs between laser pulses. It can be seen that when the particle is heated by the first laser pulse that peaks at 0 μs, the surface temperature rises rapidly and then begins to fall as the heat flux decreases. After the laser pulse, the temperature within the particle is thermalized to be uniform, where it remains until the next laser pulse occurs. Essentially, the heat transfer process can be modeled as a recurrence of laser pulses, with the uniform temperature from the previous pulse as the initial temperature of the particle for the next pulse. Since the energy delivered to the particle with each pulse is a con-stant, the temperature *rise* per pulse will be the same regardless of the particle initial temperature. The idea of pulsed laser heating is that allows chemical reaction only to occur during the pulse so that better control of the deposited film thickness can be achieved. The processing parameters must be carefully chosen so that the threshold temperature is reached but the particle temperature does not exceed the melting point. Figure 3 shows the thickness of deposited film versus time for multiple laser pulses. It can be seen that the chemical reaction does not happen until the seventh pulse, during which the surface temperature of the particle reaches the chemical reaction threshold temperature and the vapor deposition starts.

Figure 4 details the heat transfer process of the powder particle during the first laser pulse. The laser-particle interaction begins with the powder particle absorbing a significant amount of heat from the laser pulse. Figure 4(a) shows the excess surface temperature, $\theta = T - T_i$, of the particle versus dimensionless time $\tau = t/t_p$. The variation of the thermal penetration depth, $\delta_h$, versus dimensionless time $\tau$ is plotted in Fig. 4(b). When $\tau < 0$, excess surface temperature and thermal penetration depth increase with increasing heat flux at the surface. After the heat flux reaches its maximum the surface temperature starts to fall. The thermal penetration depth for $\theta_1$ sub-problem, $\delta_{h1}$ reaches the center of the particle at the time $\tau = 2.6557$, after which $\theta_1$ keeps unchanged. The thermal penetration depth for $\theta_2$ sub-problem, $\delta_{h2}$ reaches the center at the time $\tau = 4.7843$, at which point the particle has reached a nearly uniform internal temperature. The maximum surface temperature during the pulse and the uniform temperature after the pulse are 1531.7 K, 1045.2 K, respectively.

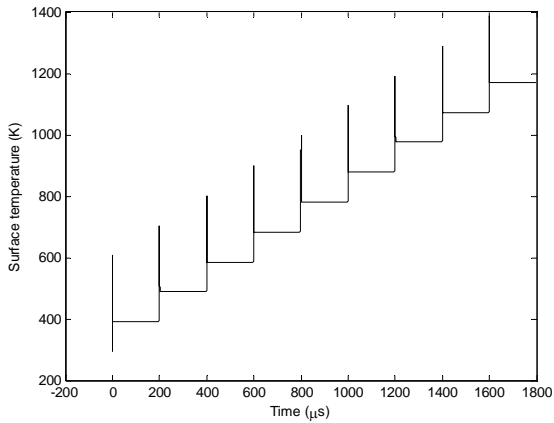

**Fig. 2.** Particle surface temperature versus time for multiple laser pulses ($J = 1688$ J/m$^2$, $t_p = 75$ ns, $r_0 = 11$ μm, $T_i = 293$ K, $p = 700$ kPa, $f = 5000$ Hz).

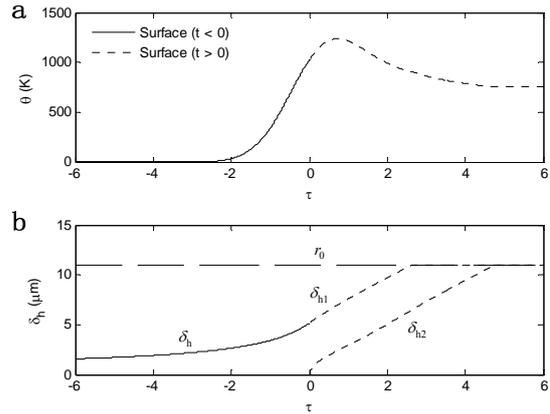

**Fig. 4.** Surface temperature and thermal penetration depth during the first laser pulse ($J = 13$ kJ/m$^2$, $t_p = 350$ ns, $r_0 = 11$ μm, $T_i = 293$ K, $p = 700$ kPa, $f = 5000$ Hz). (a) Temperatures. (b) Thermal penetration depth.

Figure 5 shows the effect of the laser pulse width, $t_p$, on the heat transfer in the powder particle. By examining Eq. (52) one can see that a decrease in $t_p$ results in an increase in heat flux, $q_0''$, if laser fluence $J$ is held constant. Thus, a shorter laser pulse width leads to quicker delivery of the laser pulse's energy, which results in an increase in surface temperature. When the pulse width is 300 ns, the maximum surface temperature during the pulse and the uniform temperature after the pulse are 1610.5 K, 1045.1 K, respectively. Since the energy input to the particle per pulse is unchanged, the final thermalized temperature is unaffected. Figure 6 shows the effect of the laser pulse width on the thickness of the deposited film. A higher surface temperature will result in a higher production rate of TiN, and a lower sticking coefficient ($\gamma_{TiN}$) when the surface temperature is above 1273 K. The higher production rate takes the priority at first and the thickness of the deposited film increases with a decrease in laser pulse width until the lower sticking coefficient comes to dominate the deposition process. After that, a thinner film will be formed with a shorter pulse width, as plotted in Fig. 6. In addition, a higher surface temperature also means that the chemical reaction happens earlier and will take longer time to stop.

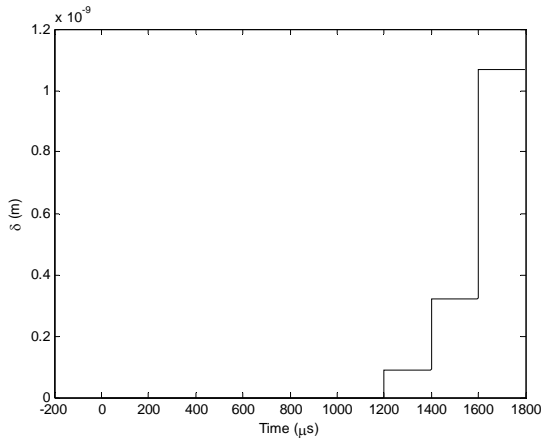

**Fig. 3.** Thickness of deposited film versus time for multiple laser pulses ($J = 1688$ J/m$^2$, $t_p = 75$ ns, $r_0 = 11$ μm, $T_i = 293$ K, $p = 700$ kPa, $f = 5000$ Hz).

The effect of the initial temperature on heat transfer and growth of thin film is shown in Figs. 7 and 8. If the initial temperature is increased, a higher surface temperature will be obtained, while the time at which the particle reaches a uniform temperature is unaffected, as well as the temperature *rise* per pulse. When the particle initial temperature is at 393 K, the maximum surface temperature during the pulse and the uniform temperature after the pulse are 1631.7 K, 1145.2 K, respectively. Figure 8 suggests that there exists a turning point when the lower sticking coefficient comes to govern the deposition process as well, after which the thickness of the deposited film grows slower with a higher surface temperature resulting from a higher particle initial temperature, i.e. fewer chemical reaction product can be stuck on the particle surface though the production rate becomes higher with a higher surface temperature.

Figure 9 shows effect of particle size on the surface temperature. It can be seen that a higher surface temperature will be reached with a smaller powder particle. When the particle ra-dius is 10 μm, the maximum surface temperature during the pulse and the uniform temperature after the pulse are 1559.1 K, 1120.5 K, respectively. This makes sense when considering that even though the input energy to all three particles is the same, the smaller particle has less mass to heat and hence reaches a higher temperature than the larger one. This also explains the higher uniform temperature of the smaller particle than that of a larger one. In addition, a smaller particle will reach a uniform temperature faster than a larger one because the heat has less material to penetrate. A turning point can also be observed in Fig. 10, before which the thickness of the deposited film grows faster with smaller particle radius and after which it goes quite the contrary since the lower sticking coefficient starts to dominate the deposition process. Finally a thicker deposited film is obtained with smaller particle radius due to the longer time for chemical reaction.

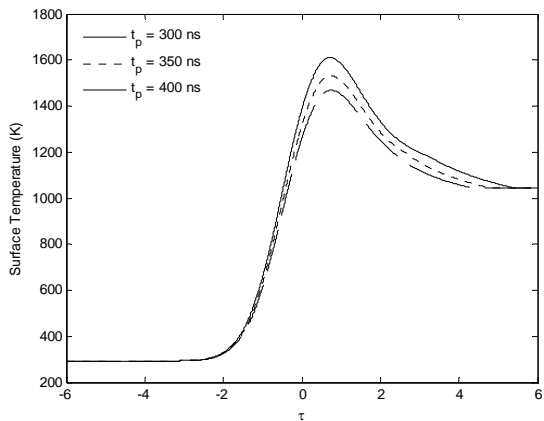

**Fig. 5.** Surface temperature during the first laser pulse for various values of laser pulse width ($J$ = 13 kJ/m$^2$, $r_0$ = 11 μm, $T_i$ = 293 K, $p$ = 700 kPa, $f$ = 5000 Hz).

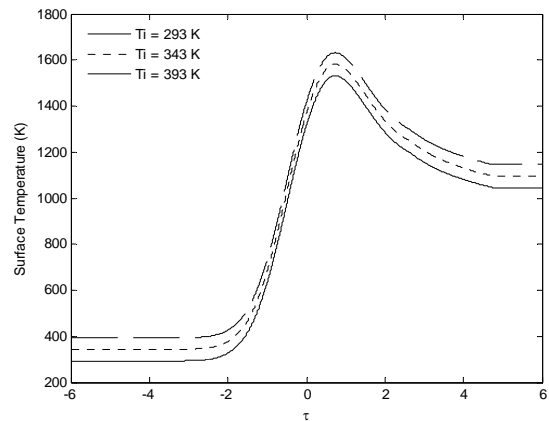

**Fig. 7.** Surface temperature during the first laser pulse for various values of particle initial temperature ($J$ = 13 kJ/m$^2$, $t_p$ = 350 ns, $r_0$ = 11 μm, $p$ = 700 kPa, $f$ = 5000 Hz).

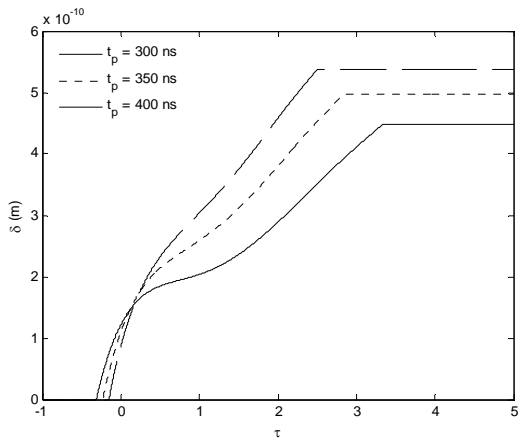

**Fig. 6.** Thickness of deposited film during the first laser pulse for various values of laser pulse width ($J$ = 13 kJ/m$^2$, $r_0$ = 11 μm, $T_i$ = 293 K, $p$ = 700 kPa, $f$ = 5000 Hz).

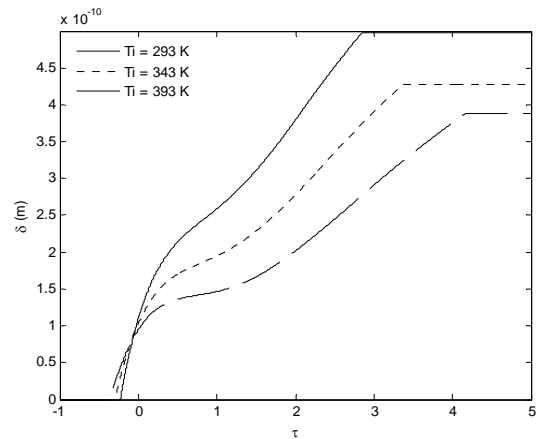

**Fig. 8.** Thickness of deposited film during the first laser pulse for various values of particle initial temperature ($J$ = 13 kJ/m$^2$, $t_p$ = 350 ns, $r_0$ = 11 μm, $p$ = 700 kPa, $f$ = 5000 Hz).

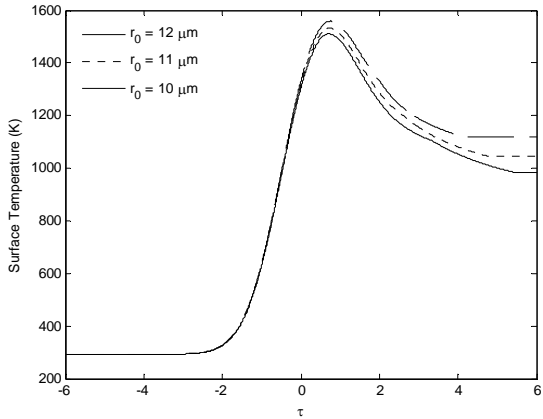

**Fig. 9.** Surface temperature during the first laser pulse for various values of particle radius ($J = 13$ kJ/m$^2$, $t_p = 350$ ns, $T_i = 293$ K, $p = 700$ kPa, $f = 5000$ Hz).

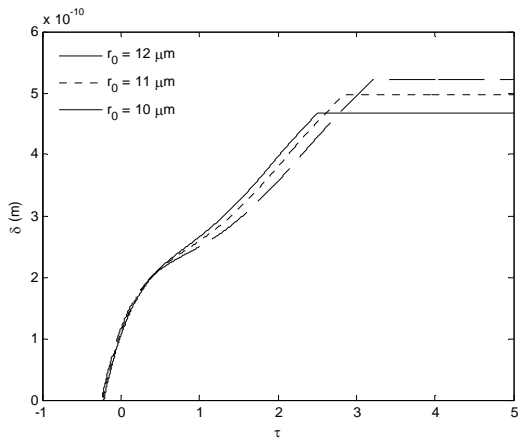

**Fig. 10.** Thickness of deposited film during the first laser pulse for various values of particle radius ($J = 13$ kJ/m$^2$, $t_p = 350$ ns, $T_i = 293$ K, $p = 700$ kPa, $f = 5000$ Hz).

Figure 11 shows the effect of laser fluence, $J$, on the particle surface temperature during the first laser pulse. Equation (52) indicates that the heat flux, $q_0''$, increases with an increase in $J$ since the laser pulse width $t_p$ is fixed. The surface temperature increases significantly with increasing laser fluence, and the particle reaches a uniform temperature at the same time for all three cases. When the laser fluence is 13 kJ/m$^2$, the maximum surface temperature during the pulse and the uniform temperature after the pulse are 1531.7 K, 1045.2 K, respectively. Figure 12 shows the thickness of deposited film during the first laser pulse for different laser fluencies. A turning point can also be observed and a thicker deposited film is obtained with higher laser fluence since the chemical reaction will last for a longer time.

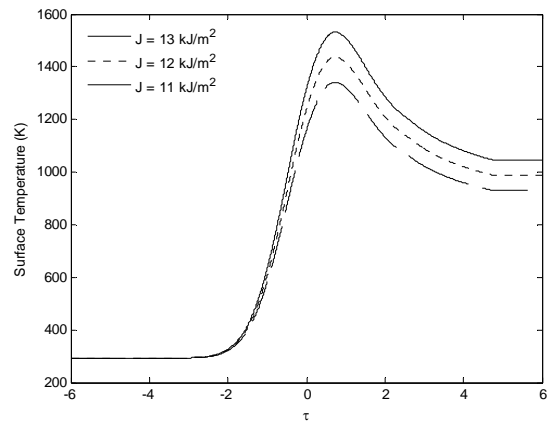

**Fig. 11.** Surface temperature during the first laser pulse for various values of laser fluence ($t_p = 350$ ns, $r_0 = 11$ μm, $T_i = 293$ K, $p = 700$ kPa, $f = 5000$ Hz).

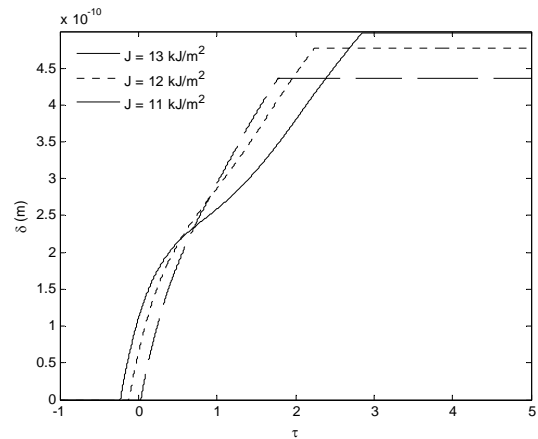

**Fig. 12.** Thickness of deposited film during the first laser pulse for various values of laser fluence ($t_p = 350$ ns, $r_0 = 11$ μm, $T_i = 293$ K, $p = 700$ kPa, $f = 5000$ Hz).

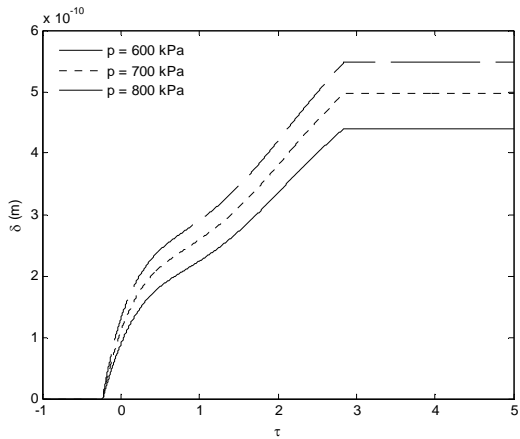

**Fig. 13.** Thickness of deposited film during the first laser pulse for various values of total pressure in the reaction chamber ($J$ = 13 kJ/m$^2$, $t_p$ = 350 ns, $r_0$ = 11 μm, $T_i$ = 293 K, $f$ = 5000 Hz).

The effect of total pressure in the reaction chamber on deposition rate is shown in Fig. 13. Increasing the total pressure in the chamber increases the global concentration of the reagent species and decreases the mass diffusivity. Finally, a thicker deposited film can be obtained due to a significant increase in $K_0$ [see Eqs. (8b) and (11)]. The deposition process starts or ends at the same time for the three cases since the heat transfer process is independent of the total pressure in the chamber.

The chemical reaction only takes place when the surface temperature is higher than the chemical reaction threshold temperature, 1173 K, and the maximum surface temperature during the pulse should not exceed the melting point of Incoloy 800, 1644 K. This significantly limits the choice of parameters, and the thickness of the deposited film has always been in the order no higher than 10$^{-9}$ m since the vapor deposition can only happen during the last few pulses. The fundamental cause for this problem is that the next pulse is launched after the particle temperature has reached to equilibrium. In real process, the next pulse can be launched before the particle temperature has become uniform after the previous pulse. However, the analytical method that was used in this paper cannot handle the situation in which the initial temperature is not uniform. Thus, our next work will be devoted to numerical solution of this problem. The particle-level model presented in this paper is an important step toward development of multiscale model of LCVI.

## 5. Conclusion

Heat and mass transfer during chemical vapor deposition on the particle surface with temporal Gaussian heat flux was investigated analytically. Since the separation time between pulses is much longer than the pulse duration, the particle temperature became uniform before the next pulse is launched. The final temperature *rise* per pulse is independent of laser pulse width and particle initial temperature since the energy delivered to the particle with each pulse is a constant. The time at which the particle reaches a uniform temperature is unaffected with the variations of particle initial temperature and laser fluence because the time necessary to penetrate the particle depends only on particle radius, $r_0$ and pulse width, $t_p$, which can be seen by examining Eqs. (22), (36) and (43). The laser fluence is the most important processing parameter, which determines the energy absorbed by a particle. A higher surface temperature, resulting from an increase in either particle initial temperature or laser fluence or a decrease in either laser pulse width or particle radius, leads to a higher production rate of TiN, and a lower sticking coefficient ($\gamma_{TiN}$) when the surface temperature is above 1273 K.

## Nomenclature

| | |
|---|---|
| $C$ | concentration [kg/m$^3$] |
| $D$ | mass diffusivity [m$^2$/s] |
| $E$ | activation energy [J/mol] |
| $k$ | thermal conductivity [W/(m·K)] |
| $K_0'$ | Arrhenius constant |
| $K_0$ | $(C_{H_2})_i(C_{N_2})_i^{1/2}K_0'$ [m/s] |
| $M$ | molecular weight [g/mol] |
| $\dot{m}$ | mass flux [kg/m$^2$] |
| $p$ | total pressure in the reaction chamber [Pa] |
| $q''$ | heat flux [W/m$^2$] |
| $q_0''$ | maximum heat flux [W/m$^2$] |
| $r_0$ | particle radius [m] |
| $R_u$ | universal gas constant (8.314 J/(mol·K)) |
| $t$ | time [s] |
| $t_p$ | half width of the laser beam pulse at $q_0''/2$ [s] |
| $T$ | temperature [K] |
| $x$ | coordinate [m] |

Greek symbols

| | |
|---|---|
| $\alpha$ | thermal diffusivity [m$^2$/s] |
| $\gamma$ | sticking coefficient |
| $\delta$ | thickness of the deposited film [m] |
| $\delta_h$ | heat penetration depth [m] |
| $\delta_m$ | mass penetration depth [m] |
| $\theta$ | temperature rise ($T$-$T_i$) [℃] |
| $\rho$ | density [kg/m$^3$] |
| $\tau$ | relative time, $t/t_p$ |

Subscripts

| | |
|---|---|
| $i$ | initial |
| $s$ | particle surface |